\documentclass[aps,pra,superscriptaddress,longbibliography,twocolumn,10pt,floatfix]{revtex4-2}

\usepackage{verbatim}
\usepackage{graphicx}
\usepackage{amsmath}
\usepackage{amssymb}
\usepackage{hyperref}
\usepackage[utf8]{inputenc}
\usepackage{mathtools}
\usepackage[english]{babel}
\usepackage{bbm}
\hypersetup{colorlinks=true, linkcolor=blue, citecolor=blue, urlcolor=blue}
\usepackage[table]{xcolor}
\usepackage{braket}
\usepackage{bbold}
\usepackage[normalem]{ulem}
\usepackage{footmisc}
\usepackage{bbm}
\usepackage{dutchcal}

\usepackage{colortbl}%
\definecolor{maroon}{cmyk}{0,0.87,0.68,0.32}

\newcommand{\bea}{\begin{eqnarray}}
\newcommand{\eea}{\end{eqnarray}}

\usepackage{pgf}
\usepackage{layouts}
\usepackage{booktabs} % For elegant tables
\usepackage{array}    % For custom column widths

\usepackage{etoolbox}
\usepackage{longtable}%
\usepackage{colortbl}%
  \newcommand{\myrowcolour}{\rowcolor[gray]{0.925}}
\usepackage{booktabs}

\begin{document}

%\linenumbers
\title{Superradiant Peak Emission Rate and Time in Quantum Emitter Arrays}

\author{Raphael Holzinger}
\email{raphael$_$holzinger$@$fas.harvard.edu}
\affiliation{Department of Physics, Harvard University, Cambridge, Massachusetts 02138, USA}

\author{Susanne F. Yelin}
\email{syelin$@$g.harvard.edu}
\affiliation{Department of Physics, Harvard University, Cambridge, Massachusetts 02138, USA}

\begin{abstract}
Determining the peak photon emission time and rate for an ensemble of $N$ quantum systems undergoing collective superradiant decay typically requires tracking the time evolution of the density operator. Generally, the dimension of the density operator grows exponentially ($\sim \! 2^N$) with the number of emitters, in the absence of any symmetries such as in Dicke superradiance with full or partial permutational symmetry. We present a detailed study of the superradiant peak emission rate and time for initially fully excited quantum emitter ensembles, for one-, two- and three-dimensional arrays in free-space and emitter chains coupled to waveguide reservoirs. For mesoscopic emitter numbers ($N\lesssim  400$) we use a second- and third-order cumulant expansion of the operator averages to track the time evolution of the system.
\end{abstract}

\maketitle
%%%%%%%%%%%%%%%%%%%%%%%%%%%%%%%%%%%%%%%%%%%%%%%%%%%%%%%%%%%%%%%%%%%%%
\section{Introduction}
\begin{figure}[ht!]
    \centering   
\includegraphics[width=0.7\columnwidth]{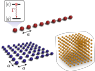}
\caption{Superradiant peak emission occurs in free-space atomic emitter ensembles, when the nearest-neighbor atomic distance $a$ is less than or comparable to the emitter resonance wavelength. In this case, the initial emission rate $R(t)$ of $N$ two-level emitters rises to a peak emission rate $R_\mathrm{peak}>N\Gamma$ at a later time. In this work we will study ensembles of two-level systems with spontaneous decay rate $\Gamma$ from the excited state ($|e\rangle$) to the ground state ($|g\rangle$) in 1D, 2D and 3D arrays embedded in a 3D photonic environment, as well as equidistant chains coupled to a waveguide (1D photonic environment).}
\label{fig:1}
\end{figure}

 Quantum emitter ensembles interacting through a shared electromagnetic reservoir can exhibit cooperative behavior in the form of superradiance~\cite{Dicke_originalpaper,gross_haroche,Anatolii_V_Andreev1980-py,haake1972quantum,PhysRevA.85.033831,PhysRevA.3.1735,benedict1996superradiance}. The case of initially fully inverted ensembles results in the cooperative enhancement of the radiative decay process and the arrival of a superradiant emission peak (see Fig.~\ref{fig:1}(b)). Accurately predicting the peak photon emission rate and time of such systems is not only a long-standing open problem in theoretical quantum optics but relevant in a wide range of research areas including superradiant lasing~\cite{haake1993superradiant,meiser2009prospects,bohnet2012steady,Kocharovsky_2017}, biochemistry~\cite{babcock2024ultraviolet}, photochemistry~\cite{monshouwer1997superradiance,doria2018photochemical}, radio astronomy~\cite{houde2018explaining,RevModPhys.95.035005}, nanomaterials~\cite{Bassani2024}, condensed matter physics~\cite{Cong_2016}, and quantum error correction~\cite{lemberger2017effect,yavuz2014superradiance}. Superradiance in the few- and many-excitation regime has been observed numerous times over the years in a broad range of platforms such as atomic gases~\cite{PhysRevA.75.033802,PhysRevA.95.043818,PhysRevLett.51.1175,PhysRevLett.117.073002,PhysRevLett.120.193601,PhysRevLett.127.243602}, solid-state
systems~\cite{PhysRevLett.115.063601,Scheibner2007,Rainò2018,Bradac2017,https://doi.org/10.1002/adfm.202102196,tiranov2023superradiance}, molecular emitters~\cite{lange2024superradiant,Kim2023}, ensembles of nuclei~\cite{Chumakov2018} and waveguide platforms~\cite{PhysRevLett.115.063601,PhysRevX.14.011020}, illustrated in Fig.~\ref{fig:1}(a).
 However, theoretically predicting the peak radiated power and time presents a significant computational challenge: tracking the time evolution of the density operator for an ensemble of $N$ quantum emitters, each undergoing collective spontaneous decay, quickly becomes intractable as $N$ increases due to the exponential growth in the dimension of the system.
 %%%%%%%%%%%%%%%%%%%%%%%%
%%%%%%%%%%%%%%%%%%%%%%%%%%%%%%%%%%%%%%%%%%%%%%%%%%%%%%%%%%%%%%%%%%%%%
The difficulty lies in the need to account for nonlinear all-to-all interactions between the two-level emitters, inherent in the superradiant decay process. Conventional methods that rely on direct simulation scale exponentially with $N$~\cite{CARMICHAEL2000417,Molmer:93,PhysRevA.45.4879}, often requiring supercomputing resources for even moderately sized emitter numbers. In the simplest case of {Dicke's} original superradiance formulation~\cite{Dicke_originalpaper}, all emitters are assumed to be indistinguishable, allowing even for analytic solutions~\cite{holzinger2025solvingdickesuperradianceanalytically,holzinger2025exactanalyticalsolutiondicke}.

In this work, we employ a second- and third-order cumulant expansion method to estimate the peak photon emission rate and time in fully inverted two-level emitter ensembles~\cite{Cumulant_Kubo,rubies2023characterizing}. The ability to provide accurate and instant predictions of peak emission properties in large-scale superradiant quantum systems, irrespective of the system configuration~\cite{mok2025universalscalinglawscorrelated}, significantly accelerates research and insights into cooperative light-matter platforms ranging from microscopic to macroscopic emitter numbers. For a fully excited initial state, we find a linear increase of the peak emission rate with $N$ for 1D arrays. In 2D and 3D arrays we find a superlinear scaling of the peak emission rate, as opposed to Dicke Superradiance or waveguide QED platforms~\cite{zhang2025robustsuperradiancespontaneousspin,lee2025exactmanybodyquantumdynamics}, which show a quadratic scaling with $N$.

\section{Theoretical description}

First, we present the theoretical framework to track the time dynamics of an ensemble of quantum emitters coupled via a common electromagnetic environment such as the free-space vacuum, a waveguide reservoir, or a dielectric medium. The system consists of $N$ two-level emitters with spontaneous decay rate $\Gamma$ and transition frequency $\omega = 2\pi c/\lambda_0$, where $\lambda_0$ is the transition wavelength between the energy levels. The emitters can be positioned arbitrarily in either free space and at the considered emitter separations, the fields emitted by each of the emitters interfere resulting in effective dipole-dipole interactions~\cite{PhysRevA.2.883}.
Using standard quantum optical techniques~\cite{charmichael_1,charmichael_2} we obtain a master equation for the internal dynamics of the emitters where the photonic part has been eliminated and the emitter density matrix $\hat{\rho}(t)$ evolves as
\begin{equation}
  \dot{\hat{\rho}}=\sum_{k=1}^{N}\Gamma_k\,
           \mathcal D[\hat O_k]\,\hat{\rho}, \qquad
  \hat O_k=\sum_{n=1}^{N} c_{k,n}\ket{g_n}\!\bra{e_n},
  \label{master}
\end{equation}
where $\mathcal D[\hat O]\hat{\rho}=\hat{O}\hat{\rho} \hat{O}^{\dagger}-\tfrac12\{\hat{O}^{\dagger}\hat{O},\hat{\rho}\}$. 
The time evolution is completely governed by the set of collective jump operators and associated decay rates $\{\hat O_k,\Gamma_k \}$ stemming from diagonalization of the (symmetric) dissipation matrix $\mathbf{\Gamma}$ with elements $\Gamma_{nm} \! = \! \frac{6 \pi \Gamma}{\omega} \mathrm{Im}[\mathbf{d}^* \cdot  \mathbf{G}(\mathbf{r}_{nm},\omega)  \cdot  \mathbf{d}]$, where $\mathbf{G}(\mathbf{r}_{nm},\omega)$ is the electromagnetic Green tensor. The spontaneous decay rate of emitter $n$ given by the diagonal element, $\Gamma_{nn}=\Gamma$ (details are presented in Appendix~\ref{supp:master}). In this work we will assume identical dipole orientations $\mathbf{d}$ for all emitters. The jump operators $\hat O_k=\sum_n c_n^{(k)} \hat{\sigma}_n$, with the coefficients obeying $\sum_{n=1}^N c_{k',n}^* c_{k,n}=\delta_{k'k}$. 
%The individual lowering operators $\sigma_n $ mediate the transition between the excited state ($|e_n\rangle$) and ground state ($|g_n\rangle$) of the $n^\mathrm{th}$ emitter. 

Eq.~(\ref{master}) describes purely dissipative dynamics and neglects any Hamiltonian terms such as coherent (dipole-dipole) exchange interaction between emitters. The influence of the coherent dipole-dipole interaction on the superradiant dynamics is small and decreases the superradiant peak emission~\cite{masson2022universality} (see the Appendix).

\vspace{1em}
The total photon emission rate on the collective transition quantifies the number of photons emitted during the (superradiant) decay process and is calculated as~\cite{robicheaux2021theoretical}
\begin{align} \label{emission}
     R(t)  = \sum_{n,m}^N \Gamma_{nm} \langle \hat \sigma^\dagger_n \hat \sigma_m \rangle = \sum_{k=1}^N \Gamma_{k} \langle  \hat {O}^\dagger_k \hat {O}_k \rangle.
\end{align}

In the limiting case of a fully excited but independent emitter ensemble, where the emitter separation (far) exceeds $\lambda_0$, the peak emission rate occurs at $t=0$ with $R_\mathrm{peak} = N \Gamma$ and is followed by an exponential decay. The limit of small emitter separations below $\lambda_0$ leads to a rapid release of photons, and to the formation of a superradiant peak with $R_\mathrm{peak} > N\Gamma$ at a later time $t>0$.

However, obtaining the actual values of the peak emission rate and the time still involved solving the early time evolution of $\hat \rho(t)$ either using the quantum master equation, whose computational cost scales exponentially, or approximate methods such as the cumulant expansion method, which scales polynomially with the number of emitters.
In the next section, we present the cumulant expansion method, for predicting the peak emission rate and time.

%%%%%%%%%%%%%%%%%%%%%%%%%%%%%%%%%%%%%%%%%%%%%%%%%%%%%%%%%%%%%%%%%%%%%
\begin{figure}[ht]
    \centering
\includegraphics[width=1\columnwidth]{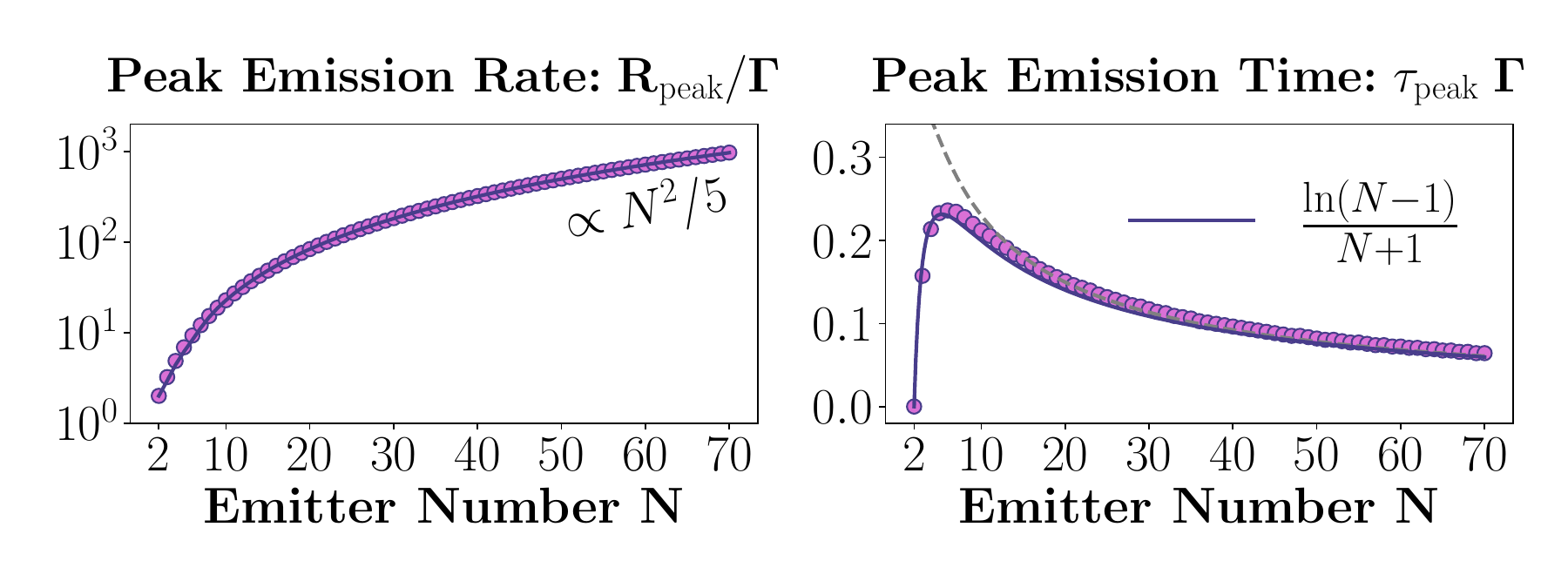}
\vspace{-2.5em}
\caption{The peak emission rate and time in Dicke superradiance based on exact numerics using the quantum master equation. The numerics are performed in the symmetric Dicke subspace were computational complexity scales linear with emitter number.
Plots of the peak emission rate and time using exact numerics based on the quantum master equation (continuous lines) as a function of $N$. The exact value of the peak emission rate converges to $\Gamma N^2/5$ at larger $N$. The peak emission time shows excellent agreement with $\ln(N \!- \!1)/[\Gamma(N+1)]$, and also shown is the literature prediction $\tau_d = \ln N / (\Gamma N)$~\cite{gross_haroche} (gray dashed line).}
    \label{fig:2}
\end{figure}
%%%%%%%%%%%%%%%%%%%%%%%%%%%%%%%%%%%%%%%%%%%%%%%%%%%%%%%%%%%%%%%%%%%%%

\subsection{Cumulant expansion method}

Starting from the Heisenberg equations of motion for spin (emitter) operators~\cite{rubies2023characterizing}, one obtains a hierarchy of equations which couple to operator correlations of ever increasing order.

In order to obtain a closed set of equations, only correlations up to a certain order are considered, while all higher-order correlations are expanded in terms of lower orders. This approximation can be computed in a systematic way using the cumulant expansion method. The joint cumulant of an $M$$^\mathrm{th}$-order correlation $\langle \hat{{O}}_1...\hat{{O}}_M \rangle$ can be expressed as an alternate sum of products of their expectation values and is compactly written as~\cite{Cumulant_Kubo}
{\begin{equation} \label{jointcumulant}
    \kappa(\hat{{O}}_1,...,\hat{{O}}_M) = \sum_P (|P|-1)!(-1)^{|P|-1} \prod_{B \in P} \Big\langle \prod_{i\in B} \hat{{O}}_i \Big\rangle
\end{equation}
where $P$ runs through the list of all partitions of $\{1, ..., M\}$, $B$ runs through the list of all blocks of the partition $P$ and $|P|$ is the number of parts in the partition.} An approximation can be made by setting $\kappa(\hat{{O}}_1,...,\hat{{O}}_M) = 0 $, which allows to express $\langle \hat{{O}}_1...\hat{{O}}_M \rangle$ with correlations of order $M-1$ and lower.
%%%%%%%%%%%%%%%%%%%%%%%%%%%%%%%%%%%%%%%%%%%%%%%%%%%%%%%%%%%%%%%%%%%%%
\begin{figure}[ht!]
    \centering   
\includegraphics[width=0.9\columnwidth]{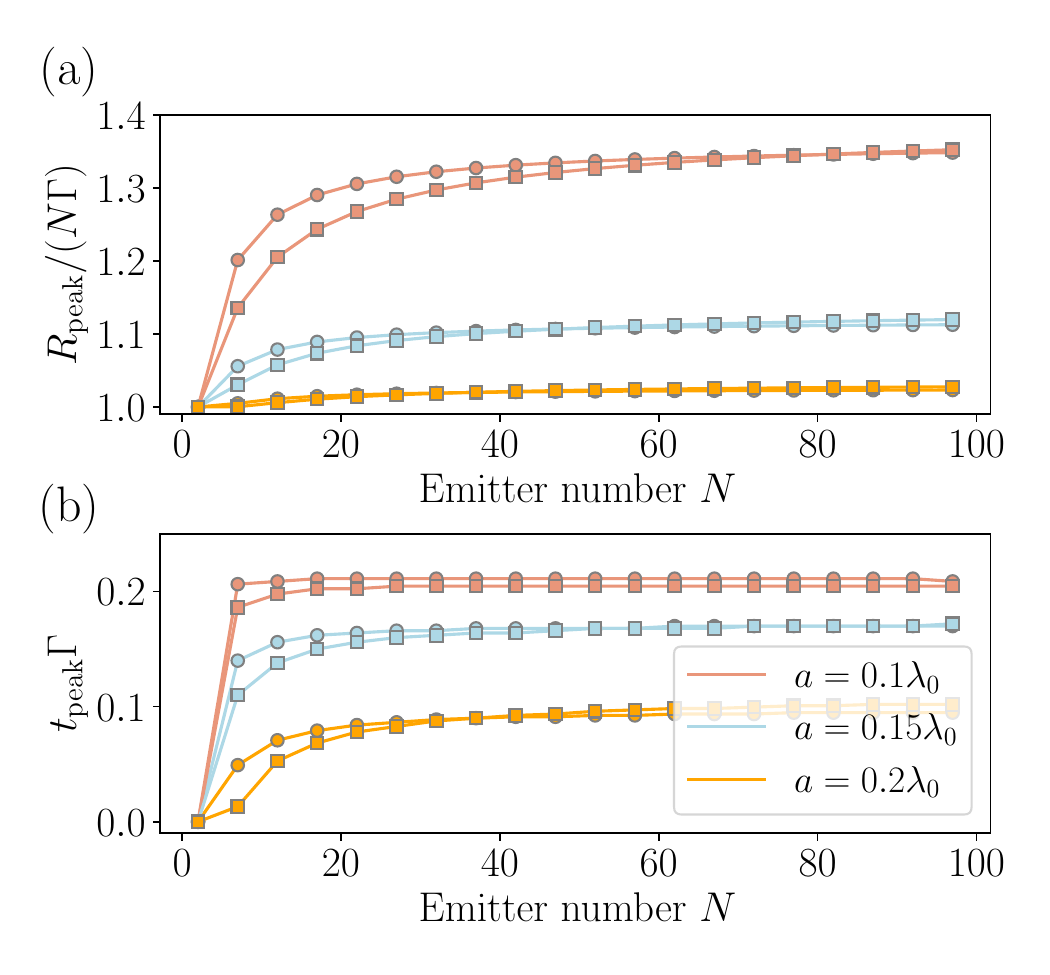}
\caption{\textbf{One-dimensional chain.} (a) The normalized peak emission rate as a function of the emitter number $N$ for various emitter spacings $a$. Shown are circular ($\circ$) and linear ($\square$) polarized dipoles, and both converge to a linear scaling with $N$. (b) The peak emission time as a function of emitter number shows a saturation as opposed to the literature result for Dicke superradiance, $t_\mathrm{peak}\Gamma = \ln(N)/(\Gamma N)$. The values are obtained with a third-order cumulant expansion.}
    \label{fig:3}
\end{figure}
%%%%%%%%%%%%%%%%%%%%%%%%%%%%%%%%%%%%%%%%%%%%%%%%%%%%%%%%%%%%%%%%%%%%%
For instance, the expansion of expectation values involving three operators reads
\begin{align}
    \langle \hat{O}_1 \hat{{O}}_2 \hat{{O}}_3 \rangle &\approx \langle \hat{O}_1  \rangle \langle \hat{O}_2 \hat{O}_3  \rangle + \langle \hat{O}_2  \rangle \langle \hat{O}_1 \hat{O}_3  \rangle \nonumber \\
    &+ \langle \hat{O}_3  \rangle \langle \hat{O}_1 \hat{O}_2  \rangle - 2 \langle \hat{O}_1  \rangle \langle \hat{{O}}_2   \rangle \langle \hat{O}_3 \rangle.
\end{align}

This introduces some error, as all higher-order correlations are neglected. In the following sections, we will show the peak emission rate and time using a second- and third-order cumulant expansion and show that the third-order exhibits excellent agreement with the master equation in Eq.~(\ref{master}) at smaller $N$.

%%%%%%%%%%%%%%%%%%%%%%%%%%%%%%%%%%%%%%%%%%%%%%

\section{Numerical results} \label{sec:superradiance}

We assume that the system is initially in the product state $|\psi_0\rangle = \bigotimes_{n=1}^N |e_n\rangle$ with its time evolution governed by Eq.~\eqref{master}.
If the decay process is superradiant, the emission peak will occur at a delayed time $t>0$, as opposed to exponential or weakly correlated decay, where the maximum emission occurs at $t=0$. To exactly find the peak emission rate and time means tracking the dissipative time dynamics and solving the superradiant decay problem with the emitter density operator $\rho$. Here, we will apply a higher-order cumulant expansion to find the peak emission rate and time for $N \lesssim 400$ emitters (second-order cumulants), $N \lesssim 200$ (third-order cumulants) and $N \lesssim 14$ (master equation).

\subsection{Dicke superradiance}
%%%%%%%%%%%%%%%%%%%%%%%%%%%%%%%%%%%%%%%%%%%%%%%%%%%%%%%%%%%%%%%%%%%%%
\begin{figure}[ht!]
    \centering   
\includegraphics[width=0.9\columnwidth]{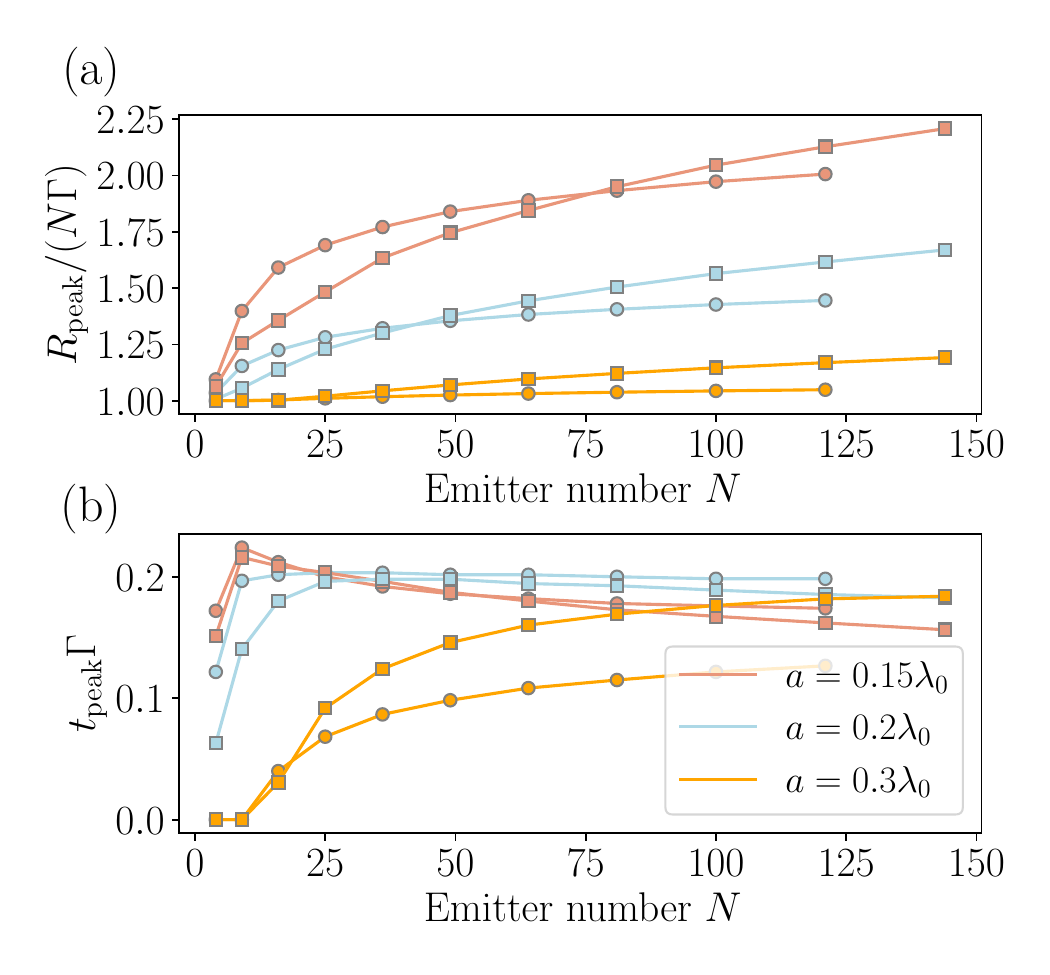}
\caption{\textbf{Two-dimensional square array.} (a) The normalized peak emission rate as a function of the emitter number $N$ for various emitter spacings $a$. Shown are circular ($\circ$) and linear ($\square$) polarized dipoles, and both converge to a superlinear scaling with $N$. (b) The peak emission time as a function of emitter number. All values are obtained with a third-order cumulant expansion.}
\label{fig:4}
\end{figure}
%%%%%%%%%%%%%%%%%%%%%%%%%%%%%%%%%%%%%%%%%%%%%%%%%%%%%%%%%%%%%%%%%%%%%
The paradigmatic example studied by Dicke~\cite{Dicke_originalpaper} of $N$ indistinguishable two-level emitters, leads to $\Gamma_{nm}=\Gamma$ for all $n,m$. Under these conditions, the maximum of Eq.~\eqref{emission} converges to $\Gamma N^2/5$ for large $N$, shown in Fig.~(\ref{fig:2}). The time of peak emission has been studied in the limit of Dicke superradiance and is historically called \textit{"delay time"}~\cite{haake1980delay,gross_haroche}, since peak emission occurs at a delayed time $t>0$. We find a peak emission time $\Gamma  \tau_\mathrm{peak} = \ln (N-1)/(N+1)$ which is close to the delay time $\ln N/(\Gamma N)$ presented in Ref.~\cite{gross_haroche}.

\subsection{One-dimensional arrays}

Quantum emitter ensembles in free-space decay collectively and superradiantly when their average nearest-neighbor separation is small enough, where the minimal spacing generally depends on the dimensionality of the ensemble~\cite{masson2022universality,masson2024dicke}. Here we assume identical (and ideal) quantum emitters in various ordered configurations, however, the model we are using applies equally to disordered ensembles. In Fig.~\ref{fig:3} peak emission rates are shown for 1D (chain) arrays, as a function of emitter number and various nearest-neighbor spacings. The exact numerics (dots) based on the quantum master equation in Eq.~\eqref{master} show good agreement with Eq.~\eqref{emission}. The corresponding peak emission times are plotted in Fig.~\ref{fig:3} with the same parameters. One can observe that the peak emission rate converges to a linear scaling with $N$ for both circular in-plane and linear out-of-plane dipole orientations as $N$ increases.

We note that Hamiltonian dynamics is not accounted for in Eq.~\eqref{master} and Eq.~\eqref{emission} (e.g. coherent dipole-dipole interactions), which generally leads to a reduction of peak emission at small spacings $a\lesssim 0.1\lambda_0$, however its influence decreases with emitter number as shown in the Appendix.

\subsection{Two-dimensional arrays}

In Fig.~\ref{fig:4}(a)-(b) we show the peak emission rates and times for two-dimensional square arrays with lattice spacing $a$ and $\sqrt{N} \times \sqrt{N}$ emitters in total. The emitters are positioned in the $xy$-plane, and are either circular ($\mathbf{d}=(1,i,0)/\sqrt{2}$) ($\circ$) or linear ($\mathbf{d}=(0,0,1)$) ($\square$) polarized. The peak emission rates show a superlinear increase with emitter number $N$, while the peak emission times show a slow linear decrease for increasing $N$.

\subsection{Three-dimensional arrays}
%%%%%%%%%%%%%%%%%%%%%%%%%%%%%%%%%%%%%%%%%%%%%%%%%%%%%%%%%%%%%%%%%%%%%
\begin{figure}[ht]
    \centering   
\includegraphics[width=0.9\columnwidth]{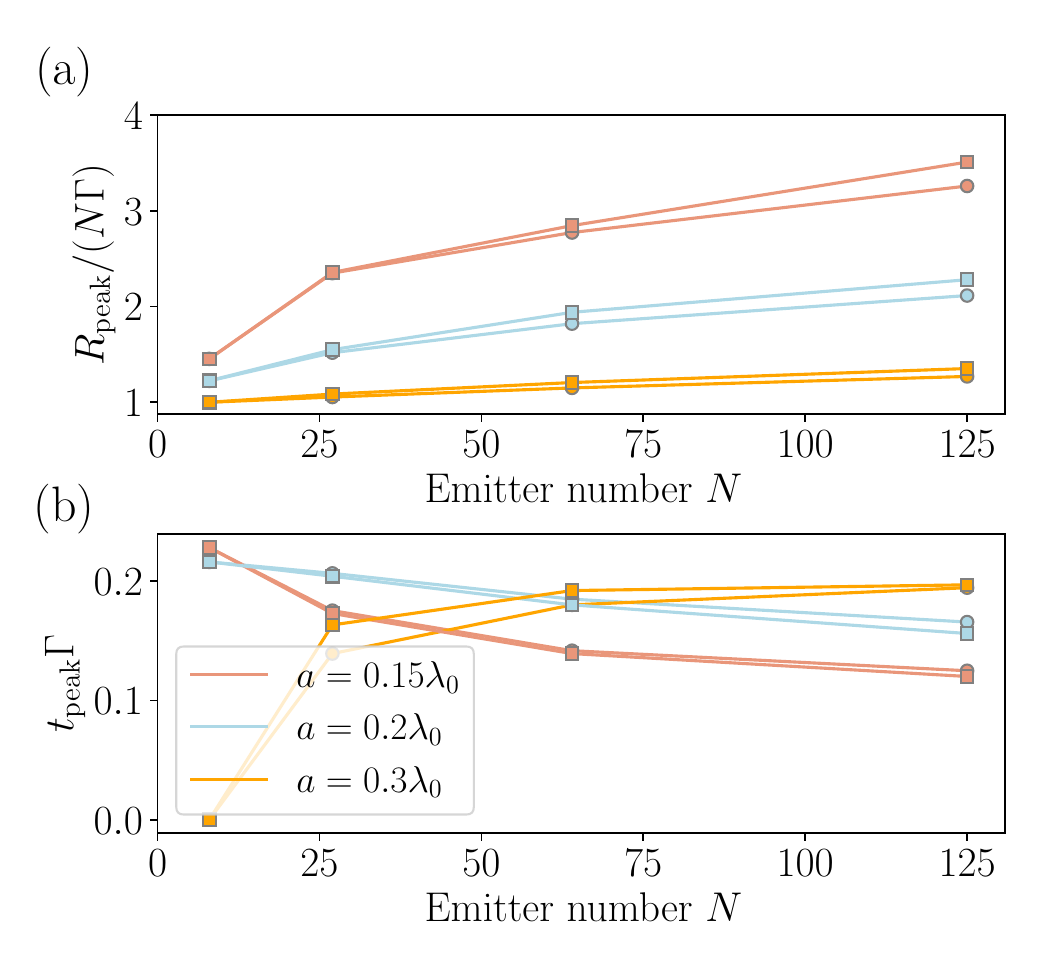}
\caption{\textbf{Three-dimensional cubic array.} (a) The normalized peak emission rate as a function of the emitter number $N$ for various emitter spacings $a$. Shown are circular ($\circ$) and linear ($\square$) polarized dipoles, and as in the case of square arrays, both converge to a superlinear scaling with $N$. (b) The peak emission time as a function of emitter number. All values are obtained with a third-order cumulant expansion.}
\label{fig:5}
\end{figure}
%%%%%%%%%%%%%%%%%%%%%%%%%%%%%%%%%%%%%%%%%%%%%%%%%%%%%%%%%%%%%%%%%%%%%
In Figs.~\ref{fig:5}(a)-(b) we show the peak emission rates and times for three-dimensional cubic arrays with lattice spacing $a$ and $\sqrt{N} \times \sqrt{N} \times \sqrt{N}$ emitters in total. The emitters are either circular ($\mathbf{d}=(1,i,0)/\sqrt{2}$) ($\circ$) or linear ($\mathbf{d}=(0,0,1)$) ($\square$) polarized and the peak emission rates exhibit, as in the square array, a superlinear increase with emitter number $N$.

\subsection{One-dimensional waveguide reservoirs}

Here, we consider an equidistant chain of emitters with lattice spacing $a$ coupled to a one-dimensional waveguide reservoir with resonant wavevector $k$. This results in a collective dissipative coupling $\Gamma_{nm}= \Gamma\cos(ka|n-m|)$~\cite{AsenjoGarcia2017,cardenas2023many}, where $\Gamma$ is the coupling rate of a single emitter to the waveguide.

\begin{figure}[ht]
    \centering   
\includegraphics[width=0.9\columnwidth]{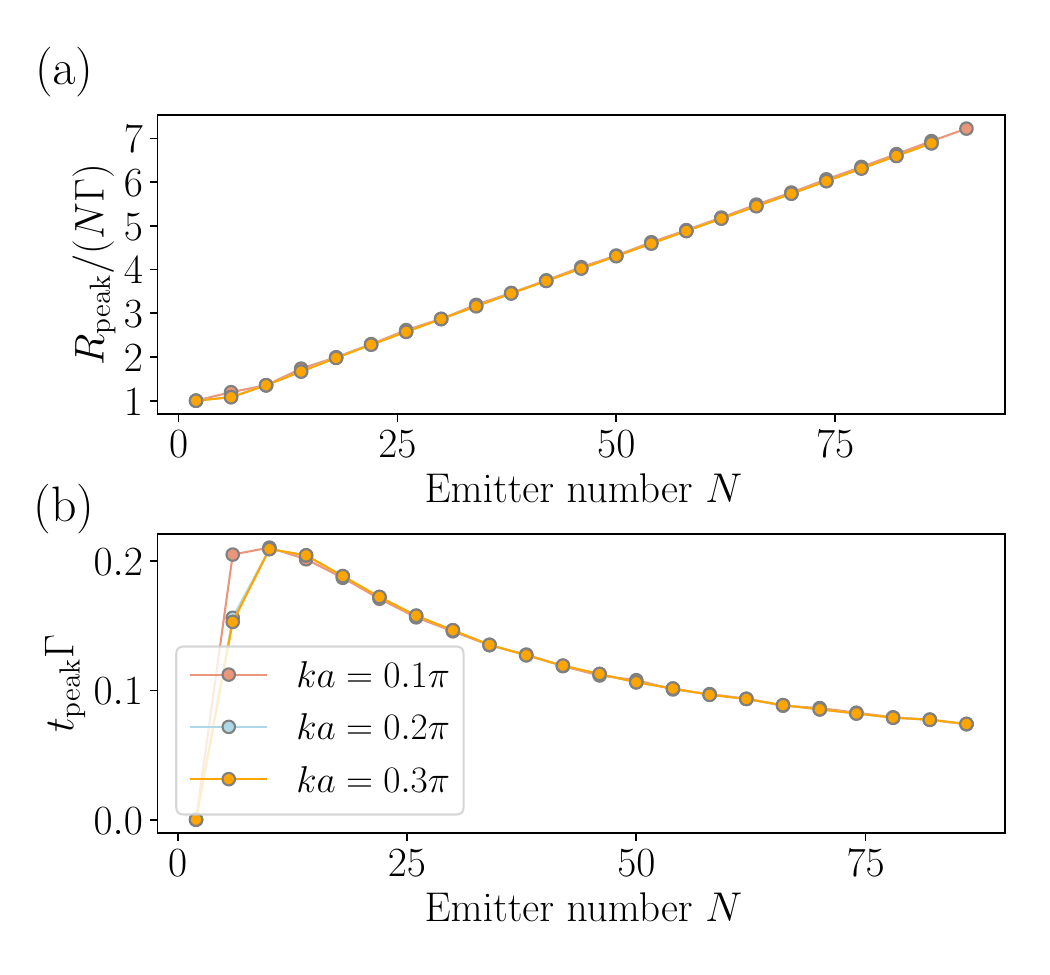}
\caption{\textbf{One-dimensional waveguide.} We consider an equidistant chain of emitters coupled to a one-dimensional waveguide with relative phase $ka$ between consecutive emitters. (a) The normalized peak emission rate as a function of the emitter number $N$ for various relative phases $ka$. The peak emission rate follows a quadratic scaling with $N$, i.e. $R_\mathrm{peak}\propto \Gamma  N^2$~\cite{lee2025exactmanybodyquantumdynamics,zhang2025robustsuperradiancespontaneousspin}. (b) The peak emission time as a function of emitter number scales as $ \ln(N)/N$ as in the case of Dicke superradiance. All values are obtained with a third-order cumulant expansion.}
\label{fig:6}
\end{figure}

In Figs.~\ref{fig:6}(a)-(b) we show the peak emission rates and times for a chain of emitters with relative phases $ka$ as a function of emitter number. The peak emission rates exhibit a quadratic scaling with $N$, i.e. $R_\mathrm{peak} \propto \Gamma N^2$~\cite{lee2025exactmanybodyquantumdynamics,zhang2025robustsuperradiancespontaneousspin} and the peak emission time is $t_\mathrm{peak} \propto \ln(N)/(\Gamma N)$, as in the case of Dicke superradiance~\cite{gross_haroche}.

%%%%%%%%%%%%%%%%%%%%%%%%%%%%%%%%%%%%%%%%%%%%%%%%%%%%%%%%%%%%%%%%%%%%%
%%%%%%%%%%%%%%%%%%%%%%%%%%%%%%%%%%%%%%%%%%%%%%%%%%%%%%%%%%%%%%%%%%%%%

\section{Conclusions and outlook}

We studied the peak photon emission rate $R_\mathrm{peak}$ and the corresponding peak time $t_\mathrm{peak}$ for fully inverted ensembles of two-level quantum emitters that decay collectively through a common reservoir. Our approach applies a third-order cumulant expansion to access mesoscopic arrays. This enables quantitative predictions of the superradiant peak emission across one-, two-, and three-dimensional free-space arrays as well as one-dimensional waveguide reservoirs governed by collective dissipation and Green-tensor-mediated couplings.

In the limit of small spacings (Dicke superradiance) we recover $R_\mathrm{peak}\!\propto\!\Gamma N^2$ and find excellent agreement for the peak time with $\Gamma t_\mathrm{peak}\!=\!\ln(N-1)/(N+1)$, close to the textbook delay-time estimate $\ln N/(\Gamma N)$. Departing from the Dicke limit, we observe distinct scaling laws that depend on array dimensionality and the photonic environment: in {free-space 1D chains}, $R_\mathrm{peak}$ grows {linearly} with $N$ for both in-plane circular and out-of-plane linear dipoles, while $t_\mathrm{peak}$ saturates with increasing $N$. In {2D square} and {3D cubic} arrays, $R_\mathrm{peak}$ exhibits a {superlinear but not quadratic} growth with $N$ and $t_\mathrm{peak}$ decreases slowly with $N$. By contrast, for {waveguide-coupled chains} the emission peak follows the Dicke-like {quadratic} scaling $R_\mathrm{peak}\propto \Gamma N^2$~\cite{lee2025exactmanybodyquantumdynamics,zhang2025robustsuperradiancespontaneousspin} and $t_\mathrm{peak}\propto \ln N/(\Gamma N)$ over a broad range of inter-emitter phases $ka$.

Methodologically, third-order cumulants closely track the master-equation benchmarks at small $N$~\cite{rubies2023characterizing,Robicheaux_cumulants} and reliably capture both $R_\mathrm{peak}$ and $t_\mathrm{peak}$ behavior in the mesoscopic regime; second-order truncation remains qualitatively correct but tends to overestimate $R_\mathrm{peak}$. In the Appendix we also quantified the role of coherent dipole–dipole exchange (Hamiltonian terms), which reduces $R_\mathrm{peak}$ at very small spacings $a\!\lesssim\!0.1\lambda_0$ but becomes progressively less important as $N$ increases, leaving the main scaling conclusions intact.

\vspace{1em}
\noindent \emph{Acknowledgments -} S.F.Y. acknowledges NSF via PHY-
2207972, the CUA PFC PHY-2317134, and QuSeC-TAQS OMA-2326787 in addition to AFOSR FA9550-24-1-0311.

%TC:ignore

%%%%%%%%%%-----------%%%%%%%%%%-----------%%%%%%%%%%-----------%%%%%%%%%% Appendixary Material

\bibliography{apssamp}

\clearpage
\pagebreak
\onecolumngrid

\appendix

\section{Quantum master equation and dipole-dipole interactions} \label{supp:master}

The system consists of $N$ two-level emitters with resonance frequency $\omega$, decay rate $\Gamma  = \omega^3 \mu^2 /(3\pi c^3 \epsilon_0 \hbar )$. By tracing out the electromagnetic field using the Born-Markov approximation~\cite{lalumiere2013input}, the emitter density matrix $\rho$ evolves in time as
\begin{align} \label{master2}
    \dot{\rho} = \sum_{n,m=1}^N \Gamma_{nm} \Big(\sigma_n \rho \sigma_m^\dagger - \frac{1}{2} \{ \sigma^\dagger_n \sigma_m, \rho \}\Big),
\end{align}
where $\sigma_n = |g_n\rangle \langle e_n|$ is the spin lowering operator for the $n^{th}$ emitter. We note that Eq.~\eqref{master2} can be written in diagonal form as
\begin{equation}
    \mathcal{L}[\rho] = \sum_{k=1}^N \Gamma_k \Big( O_k \rho O_k^\dagger - \frac{1}{2} \{O_k^\dagger O_k,\rho \} \Big),
\end{equation}
where $\Gamma_k$ are the eigenvalues of the $N\times N$ dissipation matrix with elements $\Gamma_{mn}$ and $O_k$ are the associated collective jump operators given by $O_k = \sum_{n=1}^N c_{k,n} \sigma_n$ with coefficients $c_{k,n}$ fulfilling $\sum_{k=1}^N \Gamma_k |c_{k,n}|^2= \Gamma$ and $\sum_{n=1}^N c^*_{k,n} c_{k',n} = \delta_{kk'}$. The coherent (neglected in this work) and dissipative dipole-dipole couplings between emitters $n$ and $m$ read
\begin{align} \label{3d-dipole}
   J_{nm} - \frac{i\Gamma_{nm}}{2} = -\frac{3 \pi \Gamma}{\omega} \mathbf{d}^\dagger \cdot \mathbf{G}(\mathbf{r}_{nm},\omega) \cdot \mathbf{d},
\end{align}
where $\mathbf{d}$ is the transition dipole moment matrix element, $\mathbf{r}_{nm} = \mathbf{r}_n-\mathbf{r}_m$ is the connecting vector between emitters $n$ and $m$. The Green's tensor $\mathbf{G}(\mathbf{r}_{nm},\omega)$ is the propagator of the electromagnetic
field between emitter positions $\boldsymbol{r}_n$ and $\boldsymbol{r}_m$, and reads for 3D dipole-dipole interactions
\begin{align}
\mathbf{G}(\mathbf{r}_{nm},\omega) = \frac{e^{i k_0 r_{nm}}}{4\pi k_0^2 r_{nm}^3} \bigg[ \left( k_0^2 r_{nm}^2 + ik_0 r_{nm} -1 \right) \mathbb{1}  +  \left(-k_0^2 r_{nm}^2 - 3i k_0 r_{nm} + 3 \right) \frac{\mathbf{r}_{nm} \otimes \mathbf{r}_{nm}}{r_{nm}^2} \bigg],
\end{align}
with $r_{nm} = |\mathbf{r}_{nm}|$ and $k_0 = 2\pi/\lambda_0$, where $\lambda_0$ is the wavelength of light emitted by the emitters. 
%%%%%%%%%%%%%%%%%%%%%%%%%%%%%%%

\section{Second order cumulant expansion} \label{supp:2nd-order}
\begin{figure}[ht]
    \centering   
\includegraphics[width=0.38\columnwidth]{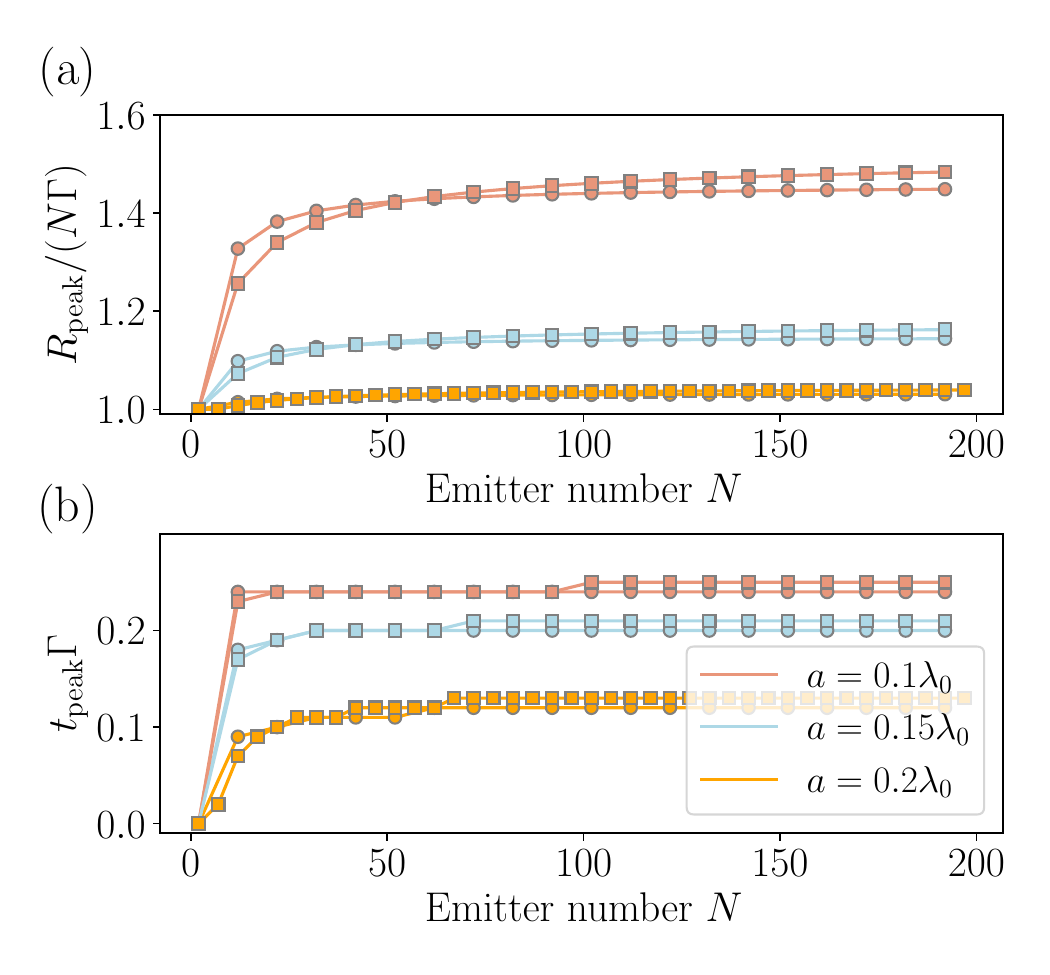}
\includegraphics[width=0.38\columnwidth]{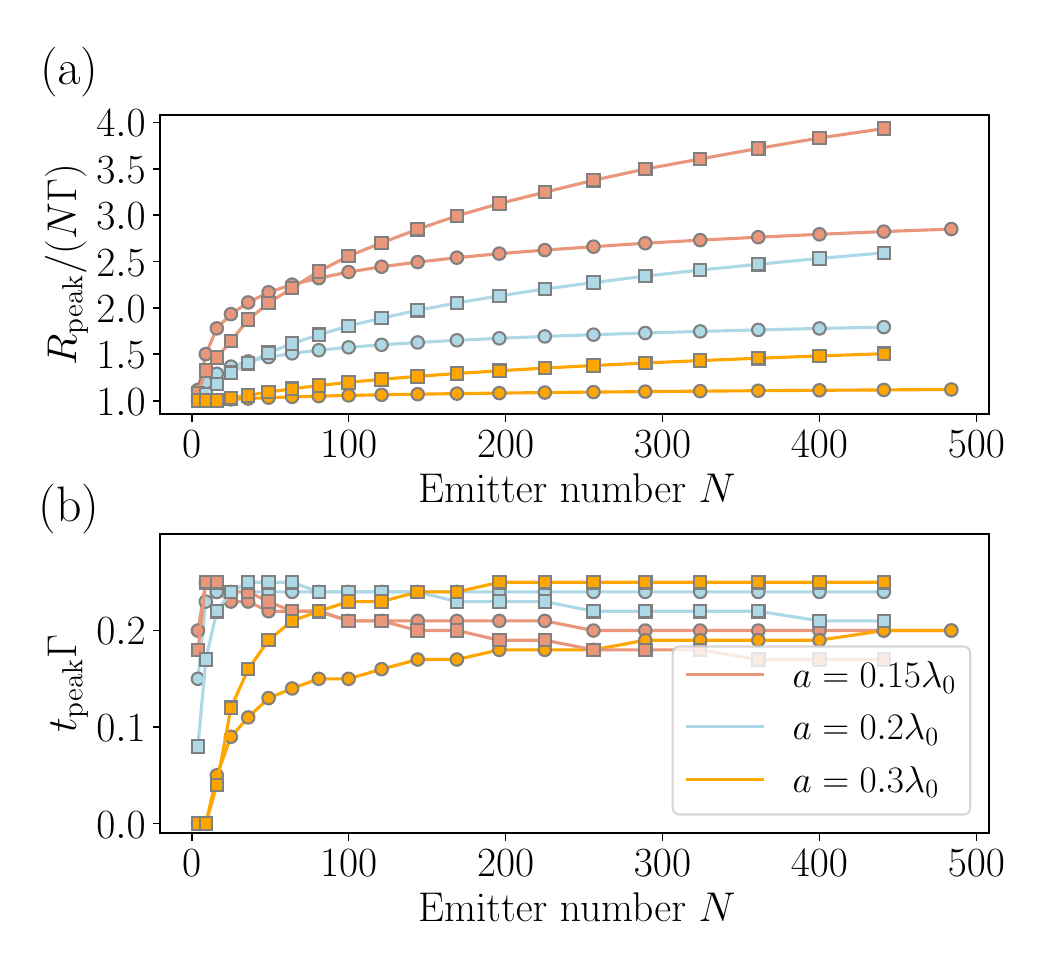}

\caption{Peak emission rates (a) and times (b) for a chain (left) and square (right) array as a function of the emitter number. The values are obtained using a second order cumulant expansion and shows an overestimation compared to the third order cumulant expansion results of the main text. Circles ($\circ$) correspond to circular ($\mathbf{d}=(1,i,0)/\sqrt{2}$) polarization and squares ($\square$) correspond to linear ($\mathbf{d}=(0,0,1)$) polarization.}
\label{fig:supp1}
\end{figure}

\begin{figure}[ht]
    \centering   
\includegraphics[width=0.38\columnwidth]{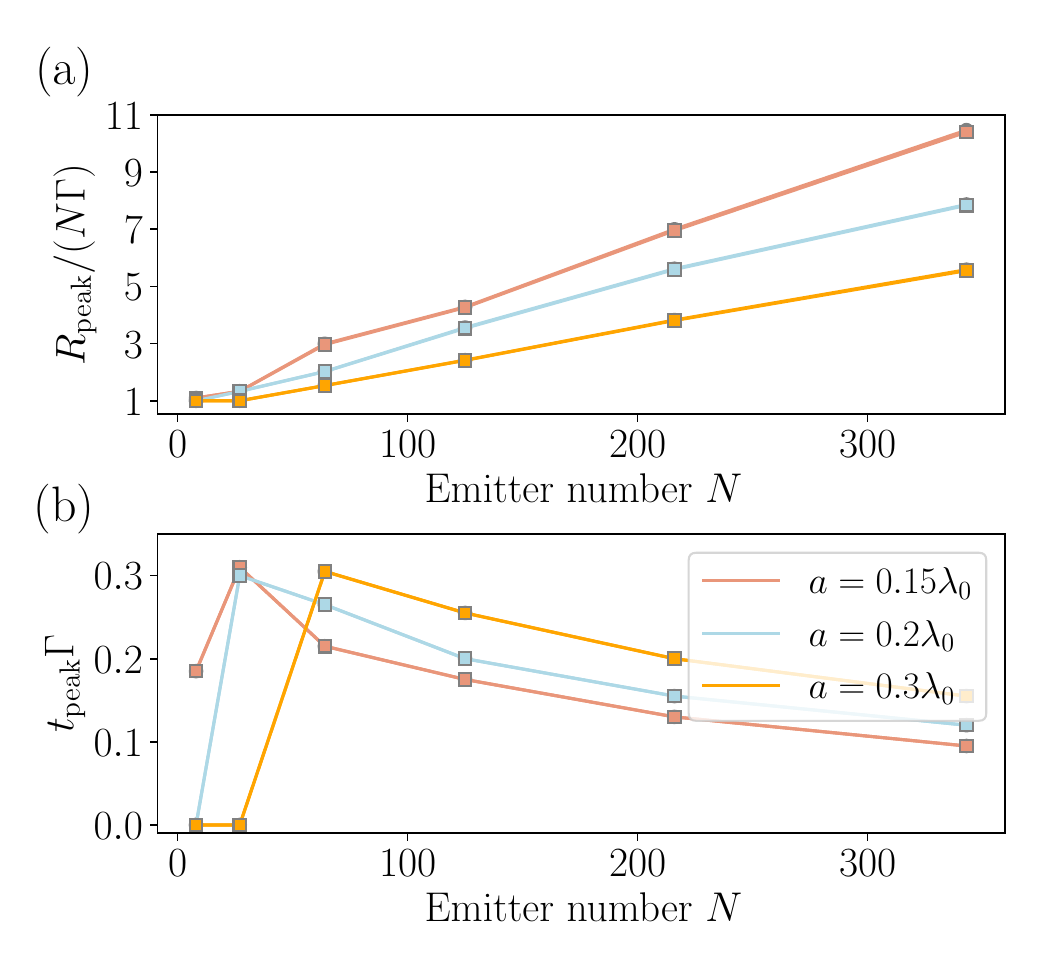}
\includegraphics[width=0.38\columnwidth]{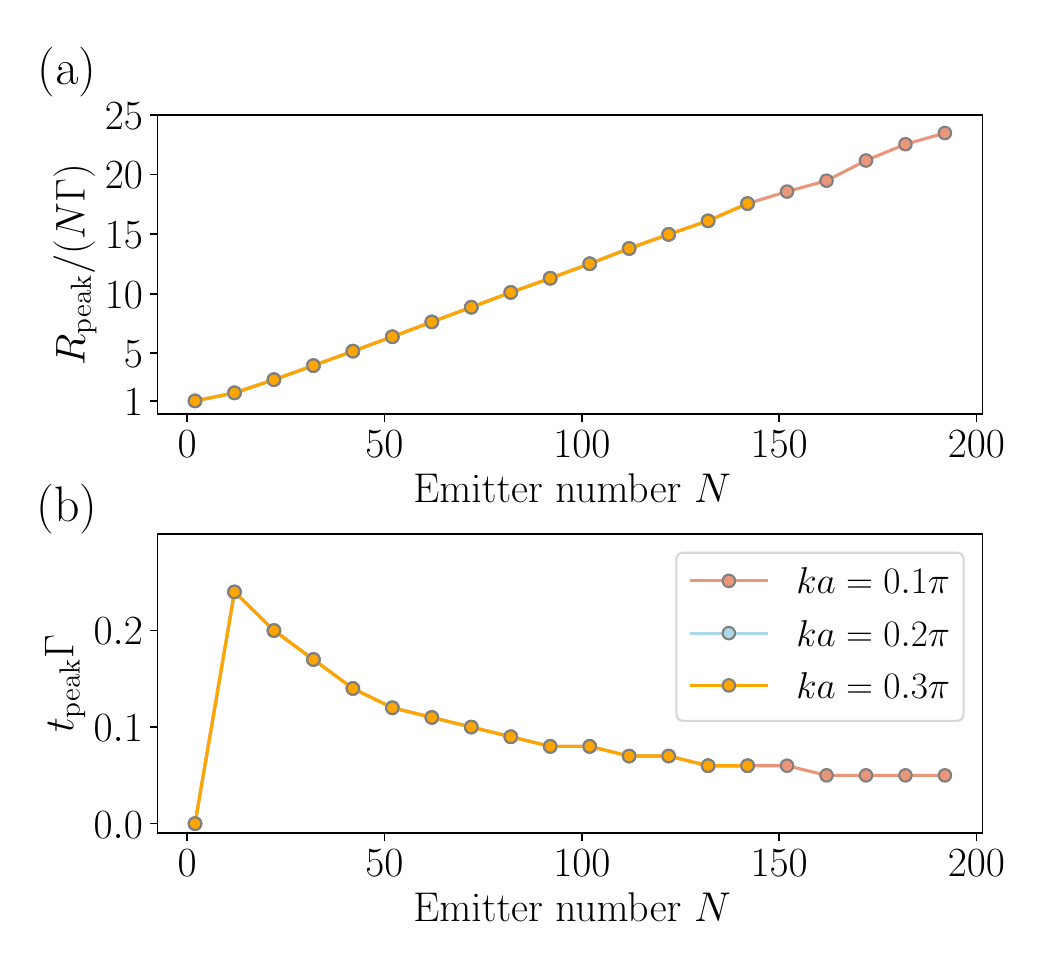}
\caption{Peak emission rates (a) and times (b) for a cubic array (left) and emitter chain coupled to a 1D waveguide (right) as a function of the emitter number. The values are obtained using a second order cumulant expansion and shows an overestimation compared to the third order cumulant expansion results of the main text. For the cubic array: Circles ($\circ$) correspond to circular ($\mathbf{d}=(1,i,0)/\sqrt{2}$) polarization and squares ($\square$) correspond to linear ($\mathbf{d}=(0,0,1)$) polarization. The waveguide coupled chain shows the same quadratic scaling of the emission rate with $N$ irrespective of the relative emitter phase $ka$~\cite{zhang2025robustsuperradiancespontaneousspin,lee2025exactmanybodyquantumdynamics}.}
\label{fig:supp2}
\end{figure}

In Figs.~(\ref{fig:supp1})-(\ref{fig:supp2}) we show the peak emission rates and times using a second order cumulant expansion.

%%%%%%%%%%%%%%%%%%%%%%%%%%%%%%%%%%%%%%%%%%%%%%%%%%%%%%%%%%%%%%%%%%%%%
%%%%%%%%%%%%%%%%%%%%%%%%%%%%%%%%%%%%%%%%%%%%%%%%%%%%%%%%%%%%%%%%%%%%%

%%%%%%%%%%%%%%%%%%%%%%%%%%%%%%%%%%%%%%%%%%%%%%%%%%%%%%%%%%%%%%%%%%%%%
%%%%%%%%%%%%%%%%%%%%%%%%%%%%%%%%%%%%%%%%%%%%%%%%%%%%%%%%%%%%%%%%%%%%%

\newpage
\section{Influence of the Hamiltonian on superradiant decay} \label{supp:Ham}
\begin{figure}[ht]
    \centering   
\includegraphics[width=0.6\columnwidth]{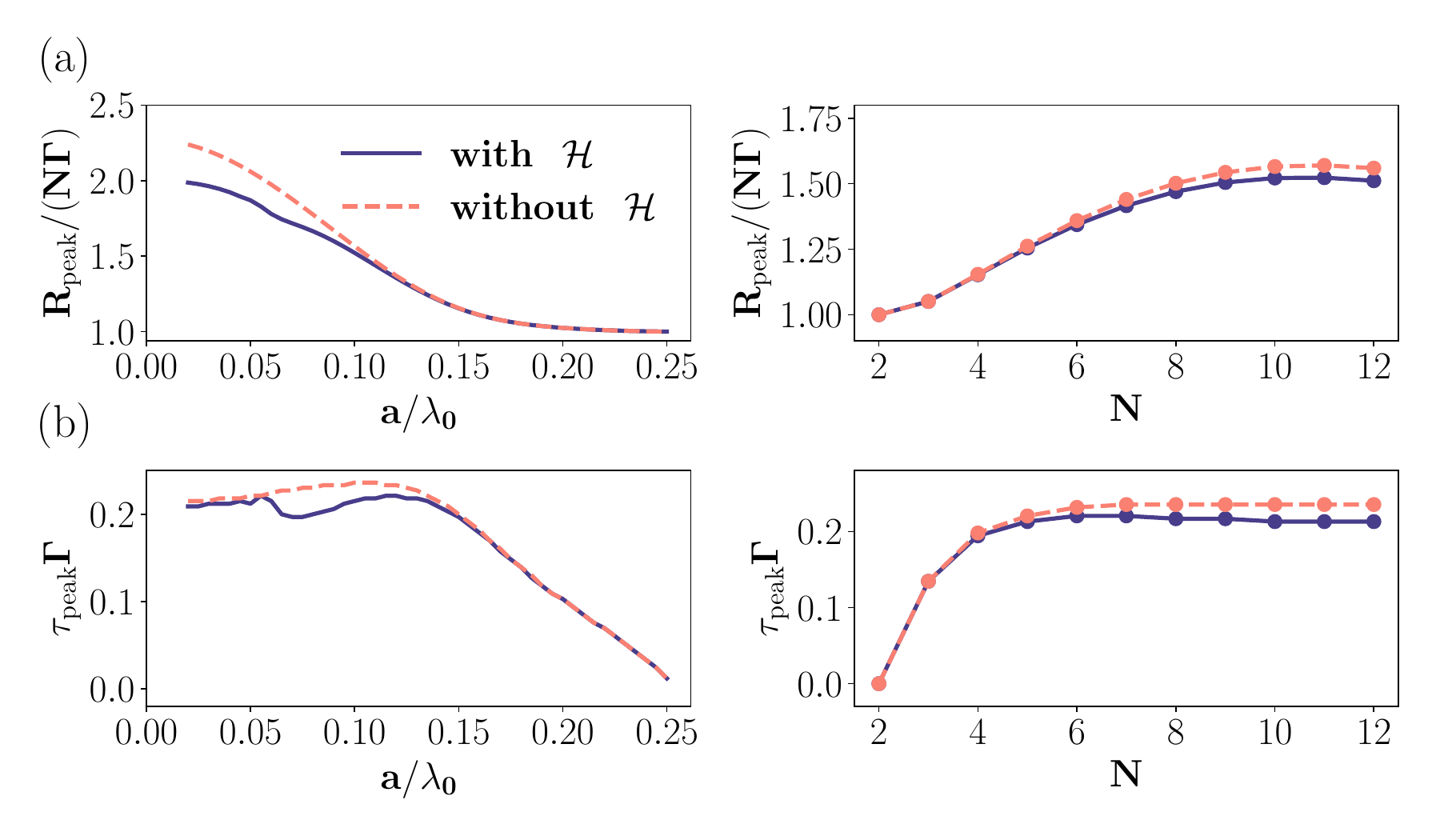}
\caption{\textbf{Free-space simulation using the quantum master equation.} Influence of the Hamiltonian on the peak emission rate and time for an emitter ring with the emitters being circularly polarized in the plane. \textbf{(a)} The peak emission rate as a function of spacing for $N=10$ (left) and as a function of $N$ for $a=0.1\lambda_0$ on the right. \textbf{(b)} The peak emission time with the same parameters as in (a).} 
    \label{fig:7}
\end{figure}

In Fig.~\ref{fig:7} the influence of the Hamiltonian dynamics in Eq.~\eqref{master2} on the peak emission properties is plotted for a ring of emitters in free space. The Hamiltonian in the rotating frame of the emitter frequency $\omega$ is given by
\begin{equation}
    \mathcal{H} =  \sum_{n,m\neq n}^N J_{nm} \sigma^\dagger_n\sigma_m,
\end{equation}
which results in coherent exchange of excitations reducing the peak emission rate at small spacings, where the coherent exchange rate becomes large.
The emitters are initially fully excited and decay either with or without the Hamiltonian term present in Eq.~(\ref{supp:master}). For smaller emitter spacings in Fig.~\ref{fig:7} the difference is more pronounced as the coherent dipole-dipole shifts $J_{nm}$ become substantial and dephases the emission process away from the symmetric Dicke states, while for the peak time a difference arises only in the interval $a\in (0.05,0.15)$.
%%%%%%%%%%%%%%%%%%%%%%%%%%%%%%%%%%%%%%%%%%%%%%%%%

\end{document}